\documentclass[aps,showpacs,prx,amssymb,amsmath,twocolumn,prr]{revtex4-2}
\usepackage{graphicx}
\usepackage{amssymb}
\usepackage{amsmath}
\usepackage{amsthm}
\usepackage{bm}
\usepackage{xcolor} 
\usepackage{array}
\usepackage{multirow}
\usepackage{tabularx}
\usepackage{booktabs}
\usepackage{textcomp}
\usepackage{mathtools}
\usepackage{hyperref}
\usepackage{qcircuit}
\usepackage{cancel}
\usepackage{comment}
\usepackage{physics}
\usepackage{romannum}
\AtBeginDocument{\pagenumbering{arabic}}

\begin{document}

\title{Impact-parameter selective Rydberg atom collision by optical tweezers}
\author{Hansub Hwang, Sunhwa Hwang, and Jaewook Ahn} 
\affiliation{Department of Physics, KAIST, Daejeon 34141, Republic of Korea } 
\author{Shuhei Yoshida}
\author{Joachim Burgd\"{o}rfer}
\affiliation{Institute for Theoretical Physics, Vienna University of Technology, Vienna, Austria, EU}
\date{\today}

\begin{abstract} \noindent
Cold collisions between two Rydberg rubidium atoms ($^{87}$Rb) are
investigated by controlling the impact parameter and collision energy. Optical
tweezers are employed to hold one atom stationary while propelling the other
to a constant velocity. After the tweezers are deactivated, both atoms are
excited to a Rydberg state by a $\pi$-pulse. After a collision, a second
$\pi$-pulse is applied. If the stationary atom does not experience a
significant momentum transfer and is de-excited to its ground state, it can
be recaptured when reactivating the tweezer. The impact parameter dependent
collision probability is extracted from the atom loss from the tweezer and
used to evaluate the collisional cross section between Rydberg atoms. 
Quantum and classical simulations of elastic two-body collisions show
good agreement with the present experimental data and provide  insights into
the critical parameter regime where quantum effects become important. 
\end{abstract}

\maketitle

\section{Introduction} \noindent
Optical tweezers are highly focused beams of light used to trap and manipulate microscopic or submicroscopic particles, leading to numerous applications across a wide range of scientific and technological fields, including biology, quantum optics, optomechanics, atomic physics, and more recently, quantum technologies~\cite{
AshkinOL1986_tweezer,GrimmAAMO2000,JohnMDoyle_molecule,SvobodaOL1994_bead,AshkinOL1987_cell_bio,quantum_level_interaction}. A
recent study has introduced a new application for optical tweezers: the use of
optical tweezers for accelerating neutral atoms~\cite{AtomTAC}. The direct
acceleration of neutral particles has remained  a challenging experimental
task, since conventional particle accelerators operate on charged
particles. Consequently, forming a neutral particle beam requires a 
neutralization step which limits the range of accessible projectiles, 
accessible energies, and degrees of beam 
focusing~\cite{DunningPRA2020,OlsonJPB1979}.
Neutral particle acceleration using optical tweezers may offer a pathway 
towards exploration of atomic and molecular collisions in a regime previously
inaccessible. 
In particular, most of low-energy collisions are studied using low-temperature
gases, making it difficult to conduct controlled experiments on molecular
collisions~\cite{KremsPCCP2008}. An experimental technique for precisely
controlling cold collisions may be applied to collisions involving a specific
number of particles rather than ensembles, or in various arbitrary collision geometries, enabling a deeper understanding of these interactions~\cite{BookofCollision, PilletPRL2017, DenschlagNP2013}. 

\begin{figure}[htbp]
\centering
\includegraphics[width=0.49\textwidth]{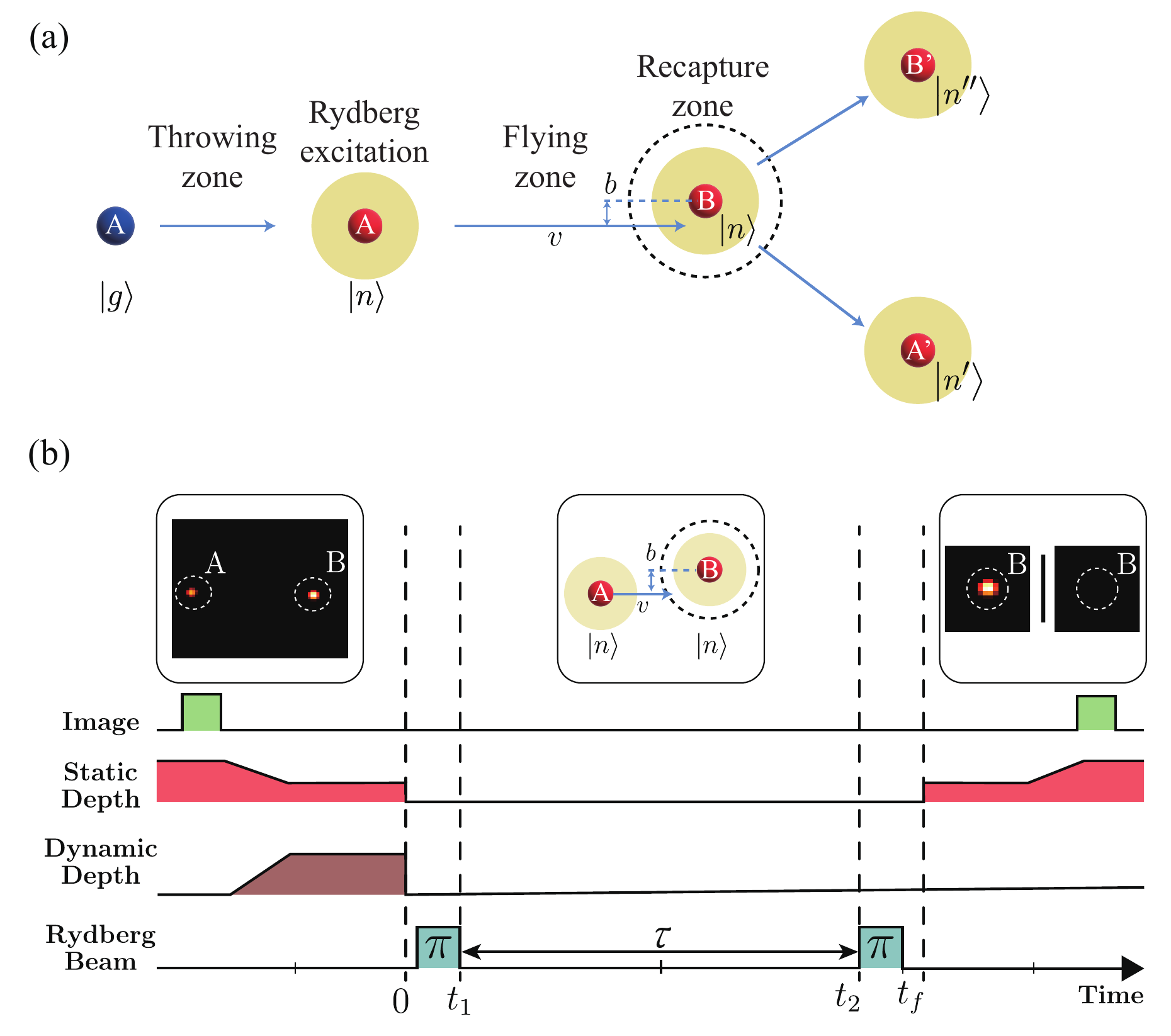}
\caption{(a) Schematic of Rydberg-atom collision experiment. (b) Experimental
  time sequence: Initially, two atoms $A$ and $B$ are captured with static
  optical dipole traps from a cloud of $^{87}\rm{Rb}$ atoms. Once the atoms
  are successfully captured, we proceed to transport atom $A$ gradually from
  the static trap to the dynamic trap for acceleration. Subsequently, after
  propelling the atom using the dynamic optical trap, we deactivate all dipole
  traps and introduce the Rydberg excitation beam to facilitate
  Rydberg-Rydberg atom collisions. Finally, we conclude the experiment by
  applying the Rydberg de-excitation beam and probing for the recapture of the
  atom $B$, to probe for the occurrence of Rydberg collisions.}
\label{Fig1}
\end{figure}

In this paper, we focus on facilitating low-energy collisions between two
Rydberg
atoms~\cite{OlsonJPB1979,Fioretti99,Zanon02,Walz04,Robicheaux05,Amthor07,Niederprum15}. We
employ optical tweezers as a means to accelerate a single rubidium ($^{87}$Rb)
atom and then observe the pairwise collision between two Rydberg-state
atoms. A schematic of the Rydberg-atom collision experiment is shown in
Fig.~\ref{Fig1}(a). A single rubidium atom (denoted as $A$), initially in the
ground state, $\ket{g}=\ket{5S_{1/2}, F = 2, m_F=2}$ (shown in blue), is
accelerated by an optical tweezer to a constant velocity $v$, while another
atom $B$, more precisely its thermal distribution, remains at rest. 
Both atoms are subsequently optically excited to the Rydberg state,
$\ket{n}=\ket{nS_{1/2}, F = 2, m_F=2}$ (shown in red) prior to their
collision. Following the collision, atom $B$ is de-excited back to $\ket{g}$
and possibly recaptured by another optical tweezer. The probability $P_B(b; n,
v)$, for the atom $B$ to be recaptured after surviving the collision, is
measured as a function of the impact parameter $b$, the relative velocity $v$,
and the principal quantum number $n$. The collisional cross section is
extracted from this recapture probability. Our results show that measured collisional cross sections agree reasonably well with predictions for classical particle collisions.

\section{Experiment} \noindent
The experimental setup encompasses a system with a magneto-optical trap for cooling $^{87}$Rb atoms, an optical tweezer configuration consisting of both static and dynamic tweezers, and an imaging system used to analyze collision events~\cite{KimNatComm2016_2dslm,LeeOE2016_3dslm,KimPRL2018_1DIsing,KimPRXq2020_3DIsing}. In the initial phase of the experiment, two rubidium atoms, cooled to a temperature of 30~$\mu$K in the magneto-optical trap (MOT), are selected from specific positions. Atom $A$ is located at coordinates $(x_A=b, 0, -z_A)$, while atom $B$ is positioned at $(0, 0, 0)$. This selection process is performed using static optical tweezers, controlled by a two-dimensional (2D) spatial light modulator (SLM, ODPDM-512 from Meadowlark Optics). Subsequently, a dynamic optical tweezer is gradually engaged to capture atom $A$. The dynamic tweezer is operated using a 2D acoustic-optic modulator (AOD, DTSxy-400-820 from AA Opto Electronics) and an arbitrary waveform generator (AWG, M4i-6622-x8 from SPECTRUM Instrument, 625 MS/s). The optical potentials and widths of the static and dynamic tweezers are $U_s=0.58(5)$~~mK, $U_d=10(1)$~mK, $d_s=0.79(4)$~$\mu$m, and $d_d=0.75(4)$~$\mu$m, respectively. These values correspond to trap frequencies of $\omega_s=67(6)$~kHz and $\omega_d=294(30)$~kHz.

The next step involves accelerating the atom $A$ along the $z$-axis to a
constant velocity in free flight~\cite{AtomTAC}. This motion is initiated by
accelerating the dynamic optical tweezer with a constant acceleration
$\ddot{z}=a$. The tweezer is deactivated at $t=0$ when it reaches the position
$z=-z_i$, releasing the atom $A$ with a velocity $v$. In this experiment, we
set the final velocities after acceleration to either $v=v_{\rm
  slow}=0.85$~m/s or $v=v_{\rm fast}=3.0(5)$~m/s. Simultaneously, the static
optical tweezers are deactivated, releasing the stationary atom $B$. Just
before the release, the temperatures of the atoms $A$ and $B$ while confined
in the optical tweezers were $T_A=150(20)$~$\mu$K and $T_B=30(4)$~$\mu$K,
respectively. For the atoms in both tweezers, these temperatures result in
position broadening of 0.06~$\mu$m for atom $A$ and 0.13~$\mu$m for atom $B$,
with corresponding velocity broadening of $\Delta v = 0.12$~m/s for $A$ around
$v$ and $\Delta v = 0.05$~m/s for $B$ around zero.

The third step in our experimental process involves excitation of the atoms to
a Rydberg state before the collision and their subsequent de-excitation after
the collision. Both atoms are excited to the same Rydberg state, chosen to be
$\ket{36S_{1/2}}$, $\ket{45S_{1/2}}$, or $\ket{53S_{1/2}}$, at $t_1 \simeq
1$~$\mu$s. After atom $A$ passes by atom $B$, both are de-excited to their
ground state at $t_2 \simeq 7 \mu$s for $v=v_{\rm fast}$ and 32~$\mu$s for
$v=v_{\rm slow}$. These particular Rydberg states are selected based on the
constraints of our frequency-locking system, which uses a dual-frequency
Fabry-Perot resonator (Stable Laser Systems, ATF-6010-4). To induce the
two-photon transition from $\ket{g}$ to the state $\ket{nS_{1/2}}$ with a
quantum defect $\delta = 3.135$, we use 780~nm and 480~nm lasers directed
perpendicular to the atoms' motion. The effective Rabi frequency for the Rydberg excitation is determined experimentally and is described by:
\begin{equation}
\Omega(\vec{r},t)=\Omega_0 e^{-\frac{x^2+y^2+(z-z_R)^2}{\sigma_R^2}}\Pi\left(\frac{t-t_{1,2}}{\pi/\Omega_0}\right),
\label{Omega}
\end{equation}
where $z_R =-4.5(3)$~$\mu$m and $\sigma_R = 6.8(1)$~$\mu$m represent the center and width of the Rydberg excitation laser, and the peak Rabi frequency is $\Omega_0/(2\pi) \simeq 1$~MHz, varying slightly depending on $n$ and $v$. The function $\Pi(x)$ is a rectangular function defined as $\Pi(x) = 1$ for $\abs{x}<1/2$ and 0 for $\abs{x}>1/2$.

In the final step, at time $t_f \simeq t_2 + 1.5$~$\mu$s, the static optical tweezer is reactivated at the original location where the atom $B$ was initially positioned. This reactivation allows to determine whether the atom $B$ is present within the recapture zone, defined by a radius of $d_B = d_s$ centered at $(0,0,0)$. A sequence of two $\pi$-pulses is applied to manipulate the electronic state of the atom at the center of the laser focus. To enhance the efficiency of de-exciting the atom $B$ with the second $\pi$-pulse, a spin-echo protocol is employed, flipping the laser phase at the midpoint $t = (t_1 + t_2)/2$. For detection, fluorescence imaging is used, based on the cycling transition $\ket{5S_{1/2}} \leftrightarrow \ket{5P_{3/2}}$. This imaging is performed with an exposure time of over 50~ms, utilizing an electron-multiplying charge-coupled device (EMCCD, Ixon Ultra 897 from Andor)~\cite{HyosubOE2019_imaging}.

Due to the velocity spread $\Delta v$ of the atom $B$, only at rest on average,
it may escape the recapture zone before the tweezer is reactivated, even
without undergoing a collision. To account for this, we alternate measurements of atom $B$ in the presence ($P_B(x_A; n, v)$) and absence ($P_B^{(0)}(n, v)$) of atom $A$. The recapture probability is then re-normalized as
\begin{equation}
P_{\rm recap}(x_A;n,v)
= \frac{P_{B}(x_A;n,v)}{P_{B}^{(0)}(n,v)}.
\label{eq:precap}
\end{equation}
Any additional atom loss can thus be attributed to collisions, and the
corresponding collision probability can be defined as
\begin{equation}
P_{\rm coll}(x_A;n,v)
= 1 - P_{\rm recap}(x_A;n,v).
\label{eq:pcol}
\end{equation}
This collision probability includes only ``hard'' elastic collisions 
with momentum transfer large enough such that atom $B$ will escape the tweezer 
irrespective on its initial position and initial thermal velocity.
The present experimental procedure is repeated about 1000 times to accumulate
the data.

\section{Theory of elastic Rydberg-Rydberg collisions } \noindent
We focus in the following primarily  on elastic scattering between Rydberg
atoms, i.e., the collision energy is preserved during the momentum transfer 
without changing the electronic state of the atoms. 
An elastic collision that prevents the recapture of atom $B$ initially at
rest in our experiment [Eq.~(\ref{eq:pcol})] requires a minimum momentum
transfer corresponding to a scattering angle $\theta > \theta_{\rm min}$ with
\begin{equation}
v \tau \sin{\frac{\theta_{\rm min}}{2}} = d_B,
\label{eq:theta_min}
\end{equation}
where $\tau = t_2 - t_1$, 
$d_B \sim 0.8$~$\mu$m the width of the tweezer,
and the propagation distance after the collision, typically, $v \tau > 10 \mu$m.
Since $d_B/ v \tau \ll 1$, 
$\theta_{\rm min} \simeq 2 d_B /(v \tau) \lesssim 0.1$~(rad). 
Therefore, atom $B$ is recaptured only at very small scattering angles
and, correspondingly, is ejected for all larger angles 
$(\theta_{\rm min} < \theta \le \pi/2)$, i.e. even for moderately ``soft''
collisions. 

It is instructive to inquire 
into the degree of classical-quantum correspondence of this scattering 
process under these kinematic constraints. 
The longitudinal de Broglie wavelength, $\lambda_{dB,\ell}$,
along the incident collision velocity can be estimated to 
be  $\lambda_{dB,\ell} = 1.5$~nm for $v = 3$~m/s and 5.4~nm for $v = 0.85$~m/s
and, thus, is much smaller that the size of the Rydberg atom
($\langle r \rangle \simeq 114$~nm for $n=36$, 186~nm for $n=45$, 
and 263~nm for $n=53$). For assessing the applicability of a classical
scattering description even more crucial is the ratio of the transverse
wavelength, $\lambda_{dB,t}$, compared to the characteristic impact parameter
$b$ determining the collision process, 
$\lambda_{dB,t}/b \ll 1$ which can be written as
\begin{equation}
p_t b \simeq \mu v \theta_{\rm qm} b \simeq L \theta_{\rm qm} \gg 1
\label{eq:ptrans}
\end{equation}
with $\mu = m/2$ the reduced mass with the atomic mass $m$ and
$p_t$ the transverse momentum.
Substituting for the classical angular momentum 
the partial-wave quantum number, $L = (\ell+1/2)\hbar$, 
Eq.~(\ref{eq:ptrans}) reduces to 
$\ell \theta \gg 1 \simeq \ell \theta_{\rm qm}$.
For classical scattering angles $\theta_{\rm min}$ [Eq.~(\ref{eq:theta_min})]
significantly larger than $\theta_{\rm qm}$, $(\theta_{\rm min} 
> \theta_{\rm qm})$,  classical-quantum correspondence should hold.
For the present case of rubidium Rydberg-Rydberg atom scattering,
the characteristic impact parameters for elastic scattering at 
the asymptotic part of the van der Waals potential are larger
than the atomic radius $\langle r \rangle$, $b \gtrsim \langle r \rangle$.
Consequently, typical values of $\ell$ are 
\begin{equation}
\ell \simeq \mu v b \gg \mu v \langle r \rangle \gg 100
\end{equation}
and, in the present case, of the order of $\sim 10^3$.
\begin{figure}
\centering
\includegraphics[width=0.4\textwidth]{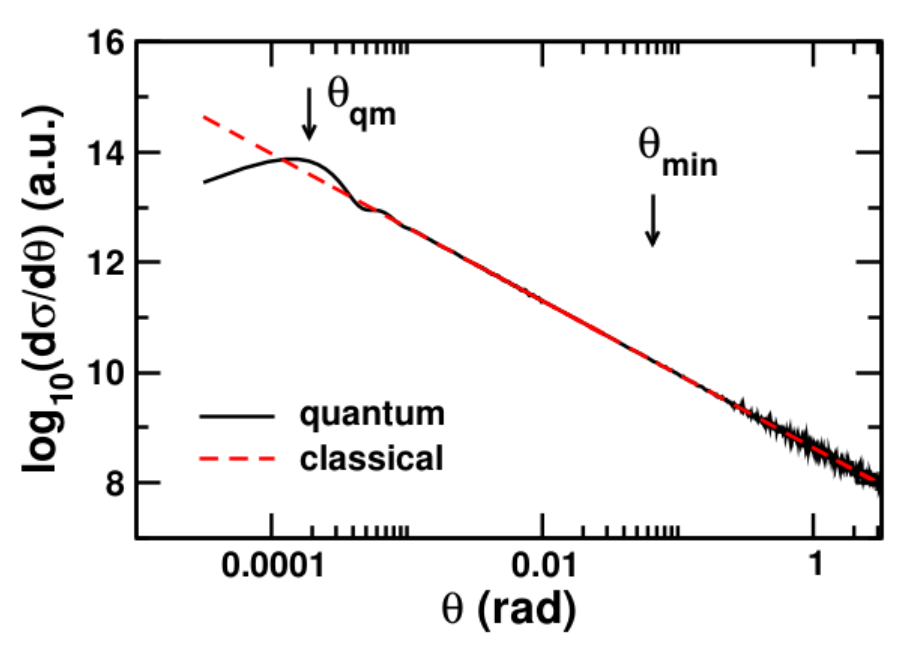}
\caption{
\label{fig:diff_cs}
Comparison between classical and quantum differential cross sections
$d\sigma/d\theta$. For scattering between rubidium Rydberg-Rydberg 
atoms with the potential $V(r) = C_6/r^6$, $n=36$ and $v=3$~m/s.
The minimum scattering angle entering the effective scattering probability,
$P_{\rm coll}$ [Eq.~(\ref{eq:pcol})], and the forward cone 
$\theta \le \theta_{\rm qm}$ where quantum effect become important
are indicated.
}
\end{figure}
For all scattering angles we find
$\theta_{\rm min} \gg \theta_{\rm qm} \simeq 10^{-3}$
and the classical limit of quantum scattering should apply.
Accordingly, the quantum scattering amplitude for scattering at 
the central potential $V(r) = C_6/r^6$ applicable to 
$r \gg \langle r \rangle$,
\begin{equation}
f(\theta) =  \frac{1}{2 i \mu v} 
\sum_\ell  (2 \ell + 1) 
  P_\ell(\cos\theta) (e^{2 i \delta_\ell} - 1)
\label{eq:sc_amp}
\end{equation}
can be evaluated in the (semi-)classical limit~\cite{Berry_Mount,brsw95}.
The partial-wave scattering phase $\delta_\ell$ for $V(r)$ can be
approximately given by~\cite{collision_book},
\begin{equation}
\delta_\ell = - \frac{3 \pi \mu^5 v^4 C_6}{16 \ell^5} \, ,
\label{eq:sc_phase}
\end{equation}
In the limit of large $\ell$ the Legendre polynomials $P_\ell$ become
fast oscillating functions and the summation in scattering amplitude 
is dominated by the terms around the stationary phase
$\ell = L_0$ satisfying 
\begin{equation}
2 \left. \frac{d \delta_\ell}{d\ell} \right|_{\ell=L_0} \pm \theta = 0 \, .
\label{eq:spa}
\end{equation}
Consequently, the scattering amplitude is
approximated~\cite{collision_book} by
\begin{equation}
f(\theta) \simeq \frac{1}{\mu v} \sqrt{
\frac{L_0}{\sin\theta}\left|\frac{dL_0}{d\theta} \right|}
e^{i (2 \delta_{L_0} - (L_0+1/2)\theta - \pi/4)} \, .
\end{equation}
Within the stationary phase approximation 
the scattering cross section becomes equivalent to that 
derived classically,
\begin{equation}
\frac{d\sigma}{d\theta}= 2 \pi \sin\theta |f(\theta)|^2 
= 2 \pi b  \left| \frac{db}{d\theta} \right|\, .
\label{eq:diff_cs}
\end{equation}
with $b = \mu v L_0$. The scattering angle $\theta$ [Eq.~(\ref{eq:spa})]
follows as 
\begin{equation}
\theta \simeq -
\int_{r_0}^\infty \frac{2 L_0}{
  r^2 \sqrt{2 \mu (E - V(r)) - L_0^2/r^2}} dr + \pi 
\label{eq:angle}
\end{equation}
with $E=\mu v^2/2$ and $r_0$ the inner turning point
satisfying $2 \mu (E - V(r_0)) - L_0^2/r_0^2 = 0$.
Expanding the integrand of Eq.~(\ref{eq:angle}) to first order in $V(r)$,
i.e., calculating the scattering angle in classical first-order
perturbation theory for the long-range portion of the van der Waals potential
yields the comparison (Fig.~\ref{fig:diff_cs}) between the quantum
differential cross section [l.h.s. of Eq.~(\ref{eq:diff_cs})]
in which $f(\theta)$ [Eq.~(\ref{eq:sc_amp})] is evaluated using
the scattering phase [Eq.~(\ref{eq:sc_phase})].
(For large scattering angle $(\theta > 0.01)$ 
the sum over $\ell$ for $\ell > 50$ is replaced by an integral). 
The classical cross section 
[r.h.s of Eq.~(\ref{eq:diff_cs})] shows excellent agreement except for the
very narrow forward cone [$\theta \lesssim \theta_{\rm qm}$, see
Eq.~(\ref{eq:ptrans})] as expected. Since in the present experiment 
$\theta_{\rm min} >  \theta_{\rm qm}$, classical two-body scattering
should suffice to the theoretically analyze the experimental data. 
We note that if the experiment could be more sensitive 
to small scattering angles, e.g., by employing smaller size tweezers or
lower the collision velocities, the wave nature of the Rydberg atom
motion would become more accessible. In the current experimental set-up, 
the threshold values $\theta_{\rm min} \simeq \theta_{\rm qm}$ are 
nearly reached for $n=36$ and $v=0.85$~m/s with a tweezer width of
$\sim 100$~nm. 

From Eq.~(\ref{eq:sc_amp}), the maximum impact parameter
$b_{\rm max}$ can be determined that is associated with the minimum scattering
angle $\theta_{\rm min}$ [Eq.~(\ref{eq:theta_min})] required to eject atom
B from the region where the optical tweezer could recapture the atom, 
\begin{equation}
b_{\rm max} = \left( \frac{15 \pi C_6 \tau}{8 m d_B v} \right)^{1/6}.
\end{equation}
Consequently, the integrated effective cross section for ``hard'' collisions
accessible by the quantity $P_{\rm coll}$ [see Eq.~(\ref{eq:pcol})] scales as
\begin{equation}
\sigma_{\rm eff}^{\rm (cl)} = 2 \pi \int_0^{b_{\rm max}} b \, db 
= \pi \left( \frac{15 \pi C_6 \tau}{8 m d_B v} \right)^{1/3}
\label{eq:cs_scaling}
\end{equation}
Given the approximate scaling $C_6 \propto n_{\rm eff}^{11}$, 
where $n_{\rm eff}$ is the effective quantum number for Rydberg atoms and 
the relation $\tau \sim 1/v$ in this setup, this cross section scales 
as $\sigma_{\rm eff}^{\rm (cl)} \propto n_{\rm eff}^{11/3} v^{-2/3}$.
More precisely, the values of $C_6$ from Ref.~\cite{ARC3.3.0}
follow a scaling of $C_6 \sim n_{\rm eff}^{11.47}$ 
(see Table~\ref{impact} for values of $C_6$, $b_{\rm max}$, 
and $\sigma_{\rm eff}^{\rm (cl)}$), 
which suggests the effective cross section scales as 
\begin{equation}
\sigma_{\rm eff}^{\rm (cl)} \propto n_{\rm eff}^{3.8} v^{-0.67}.
\label{eq:cs_scaling2}
\end{equation}
Unlike the classical total integrated cross section, its quantum counter part 
remains finite 
\begin{equation}
\sigma^{\rm (qm)}
= 2 \pi \int_0^\pi |f(\theta)|^2 \sin\theta d\theta
\simeq \frac{8 \pi}{\mu^2 v^2} \int_0^\infty L \sin^2\delta_L dL  \, .
\end{equation}
and yields for the scattering phases of the asymptotic van der Waals potential
$C_6/r^6$ the scaling law 
\begin{equation}
\sigma^{\rm (qm)} 
\simeq 5.1 \left( \frac{C_6}{v} \right)^{2/5} \,. 
\end{equation}
Correspondingly, the effective cross section for 
the same constraint of hard collisions accessed by 
$P_{\rm coll}$ is given in terms of an upper cut-off for 
$L_{\rm max} = \mu v b_{\rm max}$ by
\begin{equation}
\sigma_{\rm eff}^{\rm (qm)} 
= \frac{8 \pi}{\mu^2 v^2} \int_0^{L_{\rm max}} L \sin^2\delta_L dL  \, .
\end{equation}
Obviously, $\sigma_{\rm eff}^{\rm (qm)} < \sigma^{\rm (qm)}$.
In view of the excellent agreement for $d\sigma/d\theta$ over entire
range of accessible angles and impact parameters we can expect
$\sigma_{\rm eff}^{\rm (qm)} = \sigma_{\rm eff}^{\rm (cl)}$.

\begin{figure*}[htb!]
\centering
\includegraphics[width=0.9\textwidth]{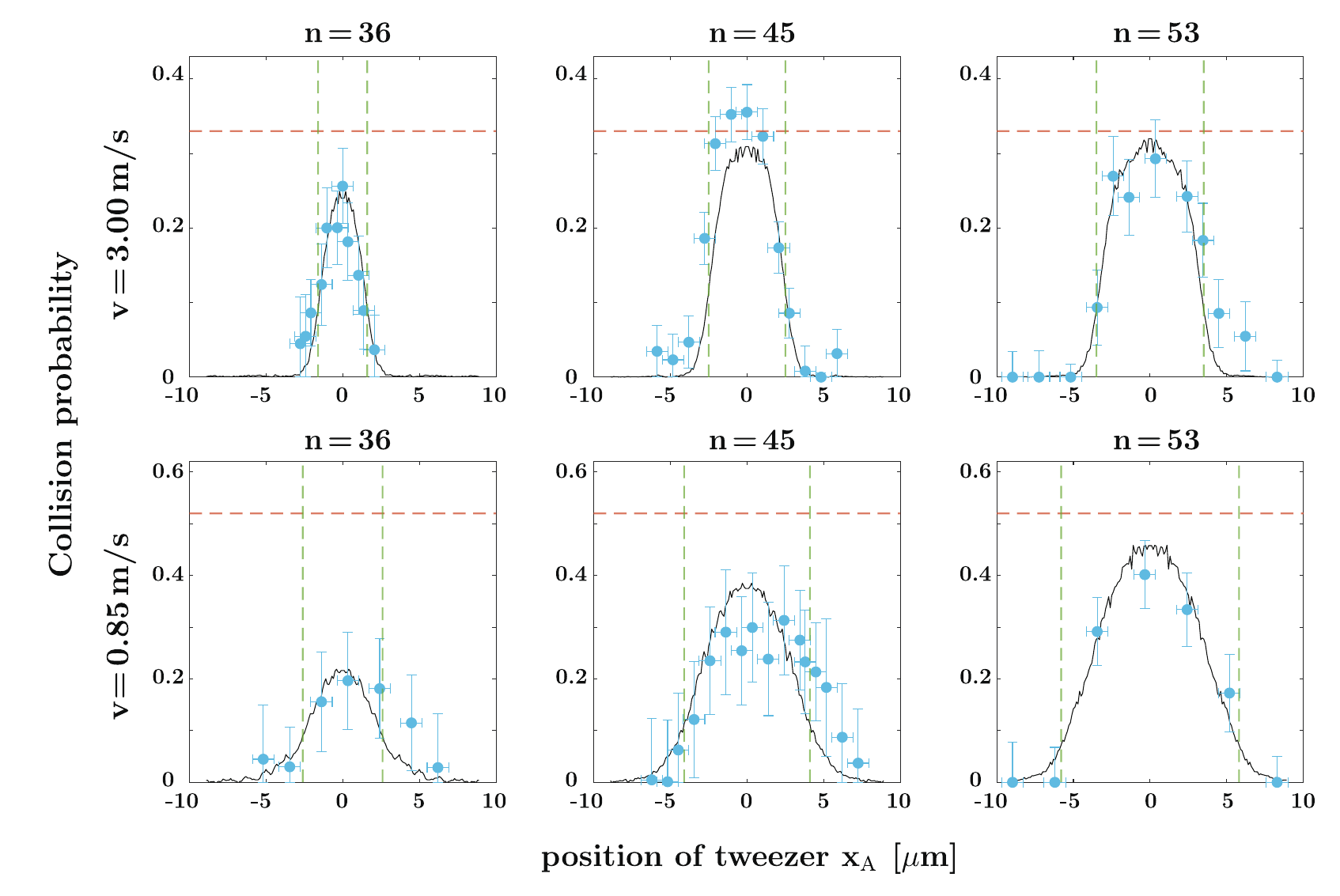}
\caption{Measured (symbols) and calculated (solid lines) collisional
probabilities for Rydberg states with $n=36$, 45, and 53 and collision
velocity of $v=3.00(6)$~m/s, and $v=0.85(6)$~m/s. Vertical dashed lines
indicate $b_{\rm max}$ in Table~\ref{impact} below which scattering is
expected ($|x_A| < b_{\rm max}$). Horizontal dashed lines indicate the photoexcitation probability $\Gamma_{AB}$.}
\label{Fig2}
\end{figure*}

\begin{table}[b]
\caption{Rydberg interaction constants~\cite{ARC3.3.0}, maximum impact
parameter $b_{\rm max}$, and effective collisional cross section 
$\sigma_{\rm eff}^{\rm (cl)}$ (Eq.~\ref{eq:cs_scaling})
for two-body Rydberg-Rydberg scattering
}
\label{impact}
\centering
\begin{ruledtabular}
\begin{tabular}{@{}c|c|cc|cc@{}}
& $C_6$ & \multicolumn{2}{c|}{$b_{\rm max}~(\mu$m)} 
& \multicolumn{2}{c}{$\sigma_{\rm eff}^{\rm (cl)}~(\mu$m$^2$)} \\
States &  (GHz/$\mu$m$^6$)  & $v_{\rm slow}$  & $v_{\rm fast}$   \quad \quad & $v_{\rm slow}$  & $v_{\rm fast} \quad \quad $  \\ \midrule
$\ket{36S_{1/2}}$ &  0.26 & 2.6 & 1.6 & 21.3 & 7.8       \\
$\ket{45S_{1/2}}$ & 4.23 & 4.1 & 2.5 & 54.1 & 19.7     \\
$\ket{53S_{1/2}}$ & 31.0 & 5.8 & 3.5 & 105.0 & 38.3     \\
\end{tabular}
\end{ruledtabular}
\end{table}

The approximate applicability of classical dynamics to the two-body scattering
allows to perform a full classical trajectory Monte Carlo (CTMC)
simulation of the experiment accounting for the thermal distribution of the 
projectile and target atoms in their respective tweezers, their electronic
(de)excitation, and their radiative decay. Field-driven electronic
(de)excitation is implemented in terms of a two-level approximation.
The initial positions and the velocities of both atoms $A$ and $B$ at $t=0$ (right after the tweezers are deactivated) are randomly
generated based on the canonical distribution of particles in an optical
tweezer. After the tweezers are turned off, while the atoms move along
straight-line trajectories, the optical Bloch equation is solved using the
position dependent Rabi frequency in Eq.~\eqref{Omega} to calculate the
excitation probability (see Table~\ref{params}). 
During the excitation phase, the interatomic distance is sufficiently 
large so that van der Waals interactions are negligible. Therefore,
the Bloch equation is solved for an isolated atom for each atom separately.
For each random realization, a Monte Carlo method is used to determine whether
the atoms are excited or not. When both atoms are excited, 
the subsequent collision dynamics are then simulated using the Hamiltonian
\begin{equation}
H = \frac{p_A^2}{2 m} + \frac{p_B^2}{2 m} 
  + \frac{C_6}{|\vec{r}_A - \vec{r}_B|^6}.
\label{eq:hamil}
\end{equation}
If neither atom is excited, they continue following straight-line trajectories without interacting. At $t=t_2$, the de-excitation probability via the second $\pi$-pulse is calculated only if atom $B$ remains within the recapture zone.
To obtain an accurate estimate, the radiative decay and dephasing in a
coherent superposition of two-level atom between the two $\pi$-pulses 
are simulated by solving the Lindblad equation~\cite{meystre}.
Since the atoms being recaptured are barely affected by Rydberg-Rydberg 
interactions, the van der Waals interaction is excluded from the simulation of
electronic dynamics. 
At the reactivation of the tweezer, $t=t_f$, the recapture probability 
of atom B, $P_{{\rm recap}, B}$ is determined by counting how many random realizations result in the atom $B$ remaining within the recapture zone and being de-excited to the ground state.

\begin{table}[h]
\caption{Parameters used for numerical simulation}
\label{params}
\centering
\begin{ruledtabular}
\begin{tabular}{cccccc}
$v$ (m/s) & $z_i$ ($\mu$m) & $t_f$ ($\mu$s) & 
  \multicolumn{3}{c}{$\Omega_0/(2\pi)$ (MHz)} \\
& & & $n=36$ & $n=45$ & $n=53$ \\
0.85 & 10.0 & 33.3 & 1.06 & 0.84 & 0.92 \\
3.0 & 14.1 & 8.2 & 1.09 & 0.88 & 0.94 \\
\end{tabular}
\end{ruledtabular}
\end{table}

\section{Result and Discussions} \noindent
Experimental and theoretical results for the collision probability as a
function of the tweezer position are presented in Fig.~\ref{Fig2}. The impact parameter dependent collision probabilities $P_{\rm coll}(x_A=b;n, v)$ are measured [Eq.~\eqref{eq:pcol}] and renormalized [Eq.~\eqref{eq:precap}], for three different Rydberg states, $\ket{36S_{1/2}}$, $\ket{45S_{1/2}}$, and $\ket{53S_{1/2}}$, and two collision velocities, $v_{\rm fast}=3.00(6)$~m/s and $v_{\rm slow}=0.85(6)$~m/s. The higher the Rydberg state ($n$) of an atom and the slower the collision velocity ($v$), the larger the position of the tweezer $d=x_A$ at which collisions occur, as expected from Eq.~\eqref{eq:cs_scaling}. The results are in a reasonable agreement with numerically simulated collision probabilities (solid lines). 

\begin{figure}[htb!]
\centering
\includegraphics[width=0.45\textwidth]{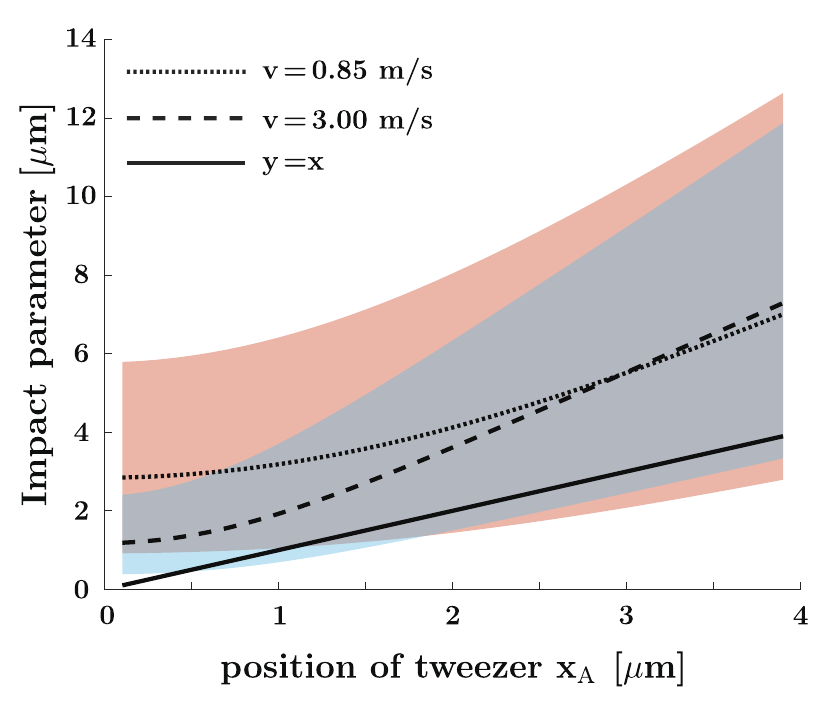}
\caption{Average impact parameter as a function of $x_A$
evaluated over randomly generated positions and velocities of atoms A and B
following the canonical ensemble in optical tweezers by the Monte Carlo simulation. For $v=3$~m/s
$x_A$ represents the average impact parameter well except at small $x_A$.
For $v=0.85$~m/s for which the thermal velocity spread 
($\sim 0.1$~m/s) becomes comparable to the average velocity
the deviations becomes rather large. For the atom A the trap frequency is 294~kHz and the temperature
150~$\mu$K centered at $\vec{r}=(x_A,0,-z_i)$ and $\vec{v} = (0,0,v)$.
For the atom B the trap frequency is 67~kHz and the temperature
30~$\mu$K centered at $\vec{r}=(0,0,0)$ and $\vec{v} = (0,0,0)$.
$z_i=10$~$\mu$m for v=0.85~m/s and 14.1~$\mu$m for $v=3$~m/s.
}
\label{fig:ave_b}
\end{figure}

The Monte Carlo simulation accounts for 
the impact parameter profiles of the two-body collisions in the current
experiment. The position of the tweezer $x_A$ serves as the representative of
the impact parameter $b$ of the two-body collision. When the actual 
impact parameter distribution is averaged over random realizations and
compared to the value $x_A$ (Fig.~\ref{fig:ave_b}), $x_A$ represents the
average impact parameter for $v=3$~m/s reasonably well except 
at small $x_A$. However, for $v=0.85$~m/s, the thermal velocity spread 
($\sim 0.1$~m/s) becomes comparable to the average velocity and
the deviations becomes rather large. Overall, $x_A$ tends to underestimate the
average impact parameter. 
The Monte Carlo simulation also displays the effect of thermal diffusion which
causes the atom $B$ to escape the recapture zone and decreases the recapture
probability, $P_B^{(0)}(x_A;n,v)$, even for large $x_A$, e.g., $P_B^{(0)}(x_A;n,v_{\rm fast}) \sim 0.56$, and  $P_B^{(0)}(x_A;n,v_{\rm slow}) \sim 0.01$.
These values correspond to the recapture probabilities without collision
$P_B^{(0)}(n,v)$. For small $x_A$ the two-body collision 
is expected to eject atom $B$ only when both atoms $A$ and $B$ are 
excited to the Rydberg state and the van der Waals interaction becomes
non-negligible. 
Therefore, the photoexcitation probability $\Gamma_{AB}$ (horizontal dashed
lines in Fig.~\ref{Fig2}) that both atoms, 
$A$ and $B$, are excited by the first $\pi$-pulse serves
as the upper boundary of the collision probability at small $x_A$.
Since the actual impact parameter $b$ can be larger than $x_A$ 
(Fig.~\ref{fig:ave_b}), the atom $B$ can be recaptured 
when $b > b_{\rm max}$ even for small $x_A$ reducing the collision probability
from $\Gamma_{AB}$.
The spontaneous decay of Rydberg states is another factor to further reduce
the collision probability.
In the current simulation the radiative lifetime of Rydberg states are 
approximated 
by $t_n = 5 \times 10^{-4} n_{\rm eff}^3$ yielding 18~$\mu$s for $n=36$,
37~$\mu$s for $n=45$, and 62~$\mu$s for $n=53$. 
For $n=53$, the lifetime and $b_{\rm max}$ become large enough such that
the collision probability in the limit of $x_A \to 0$
nearly converges to the photoexcitation probability $\Gamma_{AB}$.

Experimental effective total cross sections, 
$\sigma_{\rm eff, ex}(n,v)$, are obtained from the data for
the effective collision probabilities $P_{\rm coll}(x_A;n, v)$ via
\begin{equation}
\sigma_{\rm eff, exp}(n,v) = \frac{2 \pi}{\Gamma_{AB}} 
\int dx_A  x_A P_{\rm coll}(x_A;n,v).
\end{equation}
where we have renormalized the collision probabilities by photoexcitation
probabilities $\Gamma_{AB}$.
The resulting $\sigma_{\rm eff, exp}(n,v)$ values in Table.~\ref{crosssection}
lie below the predicted ``clean'' cross sections in Table~\ref{impact}
for two-body collisions with well-defined velocity
and impact parameter. We attribute
this mainly to the deviation of $x_A$ from the averaging over
impact parameter, as shown in Fig.~\ref{fig:ave_b}.
\begin{table}[b]
\caption{
The cross-section $\sigma_{\rm coll}(n,v)$ 
of Rydberg-atom collision data (in unit of $\mu$m$^2$).}
\label{crosssection}
\centering
\begin{ruledtabular}
\begin{tabular}{@{}cccc@{}}
States & $v = 0.85$~m/s  & $v = 3.0$~m/s   \\ \midrule
$\ket{36S_{1/2}}-\ket{36S_{1/2}}$ &  $12.1$  & $6.9$         \\
$\ket{45S_{1/2}}-\ket{45S_{1/2}}$ &  $31.4$  & $16.7$         \\
$\ket{53S_{1/2}}-\ket{53S_{1/2}}$ &  $52.5$  & $29.5$         \\ 
\end{tabular}
\end{ruledtabular}
\end{table}

\begin{figure}
\centering
\includegraphics[width=0.50\textwidth]{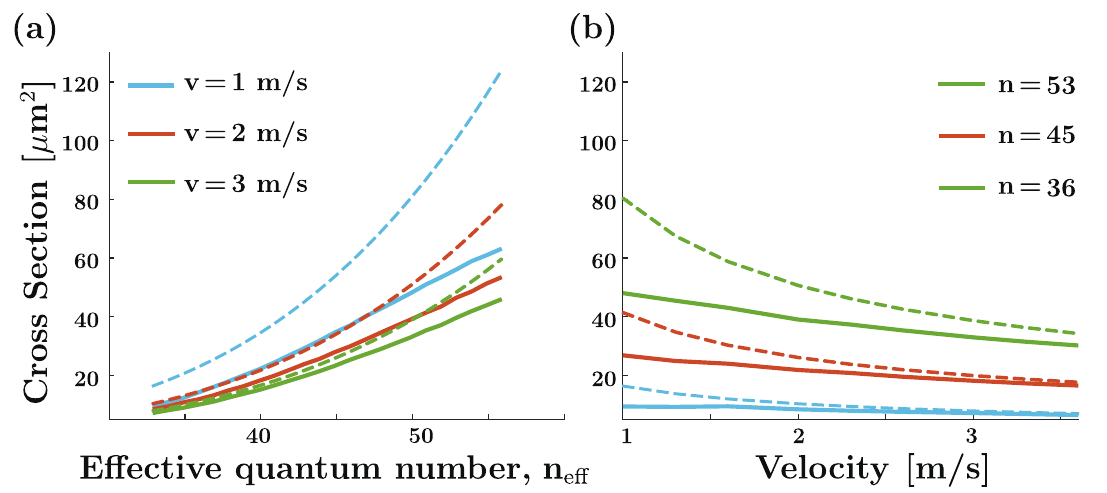}
\caption{Simulated collisional cross section for a fixed propagation distance
  (8~$\mu$m) between two $\pi$-pulses (solid lines) including thermal position
  and velocity spread. The initial position is set to $z_i = 8$~$\mu$m + $v
  \times 1.5$~$\mu$s and the Rabi frequency of $\pi$-pulses is 1~MHz. The
  total propagation time is $t_f = 2 z_i/v + 3$~$\mu$s. (a) $\sigma_{\rm
    coll}$ as a function of $n_{\rm eff}$ and (b) as a function of $v$. The
  dashed lines are the predictions of Eq.~(\ref{eq:cs_scaling}) for ``clean''
  atom Rydberg-Rydberg scattering.}
\label{fig:scaling}
\end{figure}

The numerical simulation can be further used to investigate the dependence of
the effective collisional cross section  
on effective quantum number $n_{\rm eff}$ and collision velocity $v$. 
When the propagation distance between the two $\pi$-pulses is fixed to $z_c = 16 \mu$m and, correspondingly, the collision time 
is $\tau=z_c/v$, inversely proportional to the collision velocity, 
the effective cross section is expected to follow the scaling
$\sigma_{\rm coll} \propto n_{\rm eff}^{3.8} v^{-0.67}$ 
[see Eq.~\eqref{eq:cs_scaling2}], 
being indicated in Fig~\ref{fig:scaling} by the
dashed lines. This prediction can be compared with the full
simulation allowing for the thermal spread in velocity and position.
The effective simulated cross section lies below the prediction of
the clean cross section as well.

For the dependence on the effective quantum number $n_{\rm eff}$ in 
(Fig.~\ref{fig:scaling}a), the simulation yields
$\sigma \sim n_{\rm eff}^{3.9}$ very close to the predicted scaling
while the collision velocity dependence in Fig.~\ref{fig:scaling} (b), 
is slightly off from the predicted scaling. We attribute this primarily
to the deviations of $x_A$ from the effective impact parameter in Fig.~\ref{fig:ave_b}. The deviations are most prominent for small collision velocities. As the velocity increases, the simulated results tend to converge towards 
the predicted cross sections.  


\section{Conclusion}
We have conducted an experiment of pairwise collisions of rubidium atoms in Rydberg energy states. By using optical tweezers, the initial positions and velocities of both atoms are controlled so that the cross section profiles are measured as a function of the effective impact parameter $b$ and the collision velocity $v$ for a chosen Rydberg state among $\ket{36S_{1/2}}$, $\ket{45S_{1/2}}$, and $\ket{53S_{1/2}}$. The measured results are well reproduced by a numerical simulation treating the atomic motion classically, 
and the electronic excitation as a reduced two-level system,
and accounting for thermal spread in position and velocity controlled by the
optical tweezers. We delineate the regime where quantum scattering effect will
become important.

\section{Acknowledgements}  \noindent
This work was supported by the National Research Foundation of Korea (NRF)
grant No. RS-2024-00340652, funded by the Korea government (MSIT) and by the
Austrian science fund (FWF) grant P35539-N.


\begin{thebibliography}{99}

\bibitem {AshkinOL1986_tweezer} A. Ashkin, J. M. Dziedzic, J. E. Bjorkholm, and S. Chu, ``Observation of a single-beam gradient force optical trap for dielectric particles,'' Opt. Lett. {\bf 11}, 288 (1986).
\bibitem {GrimmAAMO2000} R. Grimm, M. Weidem{\"u}ller, and Y. B. Ovchinnikov, ``Optical dipole traps for neutral atoms,''  Adv. At. Mol. Opt. Phys. {\bf 42}, 95 (2000).
\bibitem {JohnMDoyle_molecule} L. Anderegg, L. W. Cheuk, Y. Bao, S. Burchesky, W. Ketterle, K. K. Ni, and J. M. Doyle, ``An optical tweezer array of ultracold molecules,'' Science {\bf 365}, 1156 (2019).
\bibitem {SvobodaOL1994_bead} K. Svoboda and S. M. Block, ``Optical trapping of metallic Rayleigh particles,'' Opt. Lett. {\bf 19}, 930 (1994).
\bibitem {AshkinOL1987_cell_bio} A. Ashkin, J. M. Dziedzic, and T. Yamane, ``Optical trapping and manipulation of single cells using infrared laser beams,'' Nature {\bf 330}, 769 (1987).
\bibitem {quantum_level_interaction} A. M. Kaufman and K-K. Ni, ``Quantum science with optical tweezer arrays of ultracold atoms and molecules,'' Nat. Phys. {\bf 17}, 1324 (2021)

\bibitem {AtomTAC} H. Hwang, A. Byun, J. Park, S. de L{\'e}s{\'e}leuc, and J. Ahn, ``Optical tweezers throw and catch single atoms,'' Optica {\bf 10}, 401 406 (2023).


\bibitem {DunningPRA2020} S. Yoshida, J. Burgdörfer, G. Fields, R. Brienza, and F. B. Dunning, ``Giant cross sections for $L$ changing in Rydberg-Rydberg collisions,'' Phys. Rev. A {\bf 102}, 022805 (2020).

\bibitem {OlsonJPB1979} R. E. Olson, ``Rydberg-atom-Rydberg-atom ionisation cross sections,'' J. Phys. B: Atom. Mol. Phys. {\bf 12}, L109 (1979).


\bibitem{KremsPCCP2008} R. V. Krems, ``Cold controlled chemistry,'' Phys. Chem. Chem. Phys. {\bf 10}, 4079 (2008).


\bibitem {BookofCollision} S. P. Khare, ``Basics of Collisions. In: Introduction to the Theory of Collisions of Electrons with Atoms and Molecules. Physics of Atoms and Molecules,'' Springer, Boston, MA (2001).

\bibitem {PilletPRL2017} D. B. Tretyakov, I. I. Beterov, E. A. Yakshina, V. M. Entin, I. I. Ryabtsev, P. Cheinet, and P. Pillet, ``Observation of the Borromean Three-Body F{\"o}rster Resonances for Three Interacting Rb Rydberg Atoms,'' Phys. Rev. Letter {\bf 119}, 173402 (2017).

\bibitem {DenschlagNP2013} A. H{\"a}rter, A. Kr{\"u}kow, M. Dei{\ss}, B. Drews, E. Tiemann, and J. H. Denschlag, ``Population distribution of product states following three-body recombination in an ultracold atomic gas,'' Nature physics {\bf {9}}, 512-517 (2013).

\bibitem{Fioretti99} 
A. Fioretti, D. Comparat, C. Drag, T. F. Gallagher, and P. Pillet,
``Long-Range Forces between Cold Atoms,'' Phys. Rev. Lett {\bf 82} 1839 (1999)

\bibitem{Zanon02} R. A. D. S. Zanon, K. M. F. Magalhaes, A. L. de Oliveira,
and L. G. Marcassa,
``Time-resolved study of energy-transfer collisions in a sample of cold
rubidium atoms,'' Phys. Rev. A {\bf 65} 023405 (2002)

\bibitem{Walz04} A. Walz-Flannigan, J. R. Guest, J.-H. Choi, and G. Raithel,
``Cold-Rydberg-gas dynamics,'' Phys. Rev. A {\bf 69} 063405 (2004)

\bibitem{Robicheaux05} F Robicheaux, 
``Ionization due to the interaction between two Rydberg atoms,''
J. Phys. B {\bf 38} S333 (2005)

\bibitem{Amthor07} T. Amthor, M. Reetz-Lamour, S. Westermann, J. Denskat, 
and M. Weidem{\"u}ller,
``Mechanical Effect of van der Waals Interactions Observed in Real Time in an Ultracold Rydberg Gas,''
Phys. Rev. Lett. {\bf 98} 023004 (2007)

\bibitem{Niederprum15} 
T. Niederpr{\"u}m, O. Thomas, T. Manthey, T. M. Weber, and H. Ott,
``Giant Cross Section for Molecular Ion Formation in Ultracold Rydberg
Gases,'' Phys. Rev. Lett. {\bf 115} 013003 (2015)



\bibitem {KimNatComm2016_2dslm} H. Kim, W. Lee, H. Lee, H. Jo, Y. Song, and J. Ahn, ``In situ single-atom array synthesis using dynamic holographic optical tweezers,'' Nat. Commun. {\bf 7}, 13317  (2016).

\bibitem {LeeOE2016_3dslm} W. Lee, H. Kim, and J. Ahn, ``Three-dimensional rearrangement of single atoms using actively controlled optical microtraps,'' Optics Express {\bf 24} (9), 9816 (2016).

\bibitem {KimPRL2018_1DIsing} H. Kim, Y. Park, K. Kim, H.-S. Sim, and J. Ahn, ``Detailed balance of thermalization dynamics in Rydberg-atom quantum simulators,'' Phys. Rev. Lett., {\bf120}, 180502 (2018).

\bibitem {KimPRXq2020_3DIsing} M. Kim, Y. Song, J. Kim, and J. Ahn, ``Quantum Ising Hamiltonian programming in trio, quartet, and sextet qubit systems,'' PRX Quantum {\bf1}, 020323 (2020).

\bibitem {HyosubOE2019_imaging} H. Kim, M. Kim, W. Lee, and J. Ahn, ``Gerchberg-Saxton algorithm for fast and efficient atom rearrangement in optical tweezer traps,'' Optics Express {\bf27}, 2184 (2019).

\bibitem{Berry_Mount} M. V. Berry and K. E. Mount, 
``Semiclassical approximations in wave mechanics,''
Rep. Prog. Phys., {\bf 35}, 315 (1972).

\bibitem{brsw95} J. Burgd{\"o}rfer, C. O. Reinhold, J. Sternberg, and J. Wang,
``Semiclassical theory of elastic electron-atom scattering,''
Phys. Rev. A {\bf 51}, 1248 (1995).


\bibitem {collision_book} L. D. Landau and E. M. Lifshitz, Quantum Mechanics Non-Relativistic Theory (Pergamon, ed. 3, 1977).


\bibitem {ARC3.3.0} E. J. Robertson, N. Šibalić, R. M. Potvliege, M. P. A. Jones, ``ARC 3.0: An expanded Python toolbox for atomic physics calculations,'' Comp. Phys. Commun. {\bf 261}, 107814 (2021).

%
\bibitem {meystre} P. Meystre and M. Sargent,
Elements of Quantum Optics (Springer, 1998).

\end{thebibliography}
\end{document}